\documentclass[manuscript]{aastex631}

\shorttitle{Quiescent Solar Wind Regions}
\shortauthors{Short et al.}
\graphicspath{{./}{figures/}}
\usepackage[normalem]{ulem}
\begin{document}

\title{Quiescent Solar Wind Regions in the Near-Sun Environment: Properties and Radial Evolution}

\correspondingauthor{Benjamin Short}
\email{Benjamin.Short@lasp.colorado.edu}

\author[0000-0003-3945-6577]{Benjamin Short}
\affiliation{Department of Physics, University of Colorado, Boulder, CO, USA}
\affiliation{Laboratory for Atmospheric and Space Physics, University of Colorado, Boulder, CO, USA}

\author[0000-0003-1191-1558]{David M. Malaspina}
\affiliation{Laboratory for Atmospheric and Space Physics, University of Colorado, Boulder, CO, USA}
\affiliation{Astrophysical and Planetary Sciences Department, University of Colorado, Boulder, CO, USA}

\author[0000-0001-8478-5797]{Alexandros Chasapis}
\affiliation{Laboratory for Atmospheric and Space Physics, University of Colorado, Boulder, CO, USA}

\author[0000-0003-1138-652X]{Jaye L. Verniero}
\affiliation{NASA Goddard Space Flight Center, Greenbelt, MD, USA}




\begin{abstract}

Regions of magnetic field with near-radial, Parker Spiral-like geometry known as quiescent regions have been observed in Parker Solar Probe data. These regions have very low $\delta B / \langle |B| \rangle$ compared to non-quiescent solar wind at the same heliocentric distances. Quiescent regions are observed to have lower solar wind bulk speeds, lower proton temperatures, and lower proton density, consistent with properties of the slow solar wind. Inside of 15 Rs, identified quiescent regions show distinct thermal properties, having higher proton temperature anisotropies and lower parallel plasma betas compared to switchback patches observed at the same heliocentric distances. When placed on $\mathcal{R}$ vs $\beta_{\parallel p}$ plots (where $\mathcal{R}$ is the proton temperature anisotropy), quiescent region solar wind is shown to be more stable to proton cyclotron and firehose instabilities than non-quiescent solar wind at the same heliocentric distances. It is shown that quiescent regions evolve similarly to the surrounding non-quiescent solar wind, but quiescent solar wind begins at a different location in the $\mathcal{R}$ vs $\beta_{\parallel p}$ parameter space, suggesting that these regions have separate origins than the more turbulent non-quiescent solar wind. Namely, enhanced temperature anisotropies and enhanced magnetic field strength may be consistent with magnetic field lines which have undergone less magnetic field expansion compared to non-quiescent wind at the same heliocentric distances.

\end{abstract}

\keywords{Solar wind(1534) --- Interplanetary Turbulence(830) --- Space Plasmas(1544) --- Solar Magnetic Fields(1503) --- Heliosphere(711)}


\section{Introduction} \label{sec:intro}

Observed inside of 65 ${R_\odot}$ are regions of solar wind with magnetic field geometries which are closely aligned with a Parker spiral interspersed between patches of sharp magnetic deflections called “switchbacks” \citep{Bale2019,Kasper2019,deWit2020,Horbury2020a}. Quiescent regions appear to contain low levels of magnetic field fluctuations (Section \ref{subsec:delb}) and regions observed during Parker Solar Probe's (PSP) first close encounter exhibited magnetic turbulence power spectral density with a spectral break point at frequencies lower than more general solar wind \citep{deWit2020}, suggesting that the turbulence of these regions is less evolved than the surrounding solar wind. If these regions display measurable differences in their bulk properties, this difference in Alfvénic turbulence may allow us to make estimates of the role that turbulence plays in modifying the solar wind plasma parameters and through which estimate the role of turbulence in solar wind heating and acceleration. Some quiescent regions persist for hours in time series data (Section \ref{sec:data}) and quiescent regions are measured as far out as 50 ${R_\odot}$ \citep{Short2022}. This raises the following questions: how long do quiescent regions persist in the solar wind, from where do they originate, and how are they related to switchbacks if at all? Recent studies have shown that switchback patches are modulated on the scale of supergranulation networks boundaries \citep{Bale2021} at the solar surface. \citet{Bale2023} demonstrates using a Potential Field Source Surface model that switchback patches in encounter 10 are associated with open flux originating from two coronal holes. The authors of that study argue that the magnetic fields at the base of the corona undergo interchange reconnection near the edges of supergranulation network boundaries. The structure of the network boundaries then becomes embedded into the solar wind plasma which is measured by PSP at higher altitudes.

\subsection{Quiescent Region Origins} \label{subsec:introorigins}

Quiescent regions appear to be distributed between switchback patches (Figure \ref{fig:algorithm}; \citet{Bale2019,Bale2021}). Because quiescent regions are intermingled with switchback patches, it may also be the case that some quiescent regions share an association with supergranulation network boundaries. However, if quiescent regions and switchback patches share this supergranulation network association, why then is it that they possess different turbulence properties such as differing spectral break points \citep{deWit2020} and different magnetic field fluctuation amplitudes (Figure \ref{fig:delb})? Results from \citet{deWit2020} suggest that quiescent regions have properties of less evolved solar wind. These statements can be used to generate hypotheses: hypothesis 1 says that quiescent regions and switchback patches originate from similar locations on the solar surface and at the same altitude. If this hypothesis is true then the location from which quiescent regions originate must have properties that inhibit the development of large scale magnetic field fluctuations. Hypotheses 2 and 3 suggest that switchback patches do not share an origin location with switchback patches. Because many quiescent regions are distributed between switchback patches, it is expected that at least some quiescent regions originate along the similar field lines that switchback patches do, but it is not necessary that they originate from the same altitude. Hypothesis 2 says that quiescent regions originate from deeper within the corona than switchback patches do and the pristine properties of the younger solar wind pass through the magnetic field fluctuations induced by interchange reconnection or other mechanisms. Hypothesis 3 states that quiescent regions originate from higher altitudes than more turbulent solar wind regions like switchback patches and develop in-situ.

If hypothesis 1 is correct, it indicates that the structure of the supergranulation network boundaries and events on the solar surface such as interchange reconnection \citep{Bale2023} lead to the appearance of solar wind regions which take on distinct characteristics. Hypothesis 2 suggests that there are lanes through which pristine plasma from low in the corona can penetrate the turbulence of the lower corona without becoming unstable to switchback growth. Hypothesis 3 necessitates a mechanism by which distinct turbulent properties develop between quiescent and non-quiescent solar wind, only for those differences to be filled back in by the time the solar wind reaches 1 AU. This study seeks to determine the properties of quiescent regions measured at a range of solar altitudes and distinguish between these possibilities. 

Section \ref{sec:data} details the data used in this study and how they are selected. It also details the data processing algorithm used to identify quiescent regions. Section \ref{sec:results} details the results of this study. Section \ref{sec:discussion} discusses the possible interpretations of the results and Section \ref{sec:conclusion} summarizes the interpretations and compares them to hypotheses above.

\section{Data Set and Processing} \label{sec:data}

\subsection{Data Products} \label{subsec:data}

This study utilizes data collected from the FIELDS instrument suite \citep{Bale2016} as well as the Solar Wind Electron, Alphas, and Protons (SWEAP) \citep{Kasper2016,Livi2021} instrument suite. From FIELDS we use survey data from the fluxgate magnetometers. From SWEAP we use measurements from the Solar Probe Cup (SPC) \citep{Case2020} and the Solar Probe ANalyzers Ion (SPAN-Ion) \citep{Livi2021} sensors. This study uses data from encounters 1 through 16 and covers radial distances of 13.3 $R_{\odot}$ to 65 $R_{\odot}$. The fluxgate magnetometers are located on a 3.5 m boom on the rear of the spacecraft opposite the Thermal Protection Shield (TPS). While these magnetometers operate at a cadence of roughly 293 Sa/s \citep{Bale2016} this study uses a downsampled cadence of roughly 4.53 Sa/s which is available in the L2 public data products. This study also uses particle distributions and derivative quantities from the SWEAP instruments for protons {\citep{Kasper2016,Livi2021}}. Using the publicly available L3 SPAN-I moments, we calculate the proton temperatures, densities, bulk velocities, plasma betas, Alfvén velocities, and temperature anisotropies.

Quiescent regions have been defined in terms of the normalized deflection angle from a theoretical Parker Spiral \citep{deWit2020}. This deflection angle, $z$, is defined as

\begin{displaymath}
    z = \frac{1}{2} (1-\cos(\theta_{def})) = \frac{1}{2} (1-\mathbf{\hat{b}_{meas}} \cdot \mathbf{\hat{b}_{Park}})
\end{displaymath}

\noindent where $\theta_{def}$ is the deflection angle of the measured magnetic field away from a theoretical Parker Spiral and its cosine is the dot product of the measured magnetic field unit vector $\mathbf{\hat{b}_{meas}}$ with the Parker Spiral unit vector  $\mathbf{\hat{b}_{Park}}$. The unit vector $\mathbf{\hat{b}_{meas}}$ is measured by the fluxgate magnetometers and reported in RTN coordinates. The unit vector for the Parker Spiral is derived using a combination of the radial component of the solar wind velocity measurements from SPAN-I and from SPC.

\subsection{Quiescent Region Identification} \label{subsec:algorithm}

This study seeks times where the solar wind matches the previously used \citep{deWit2020} $z$ threshold of $z < 0.05$ (or $z > 0.95$ when PSP crosses the Heliospheric Current Sheet and the global field polarity changes sign) but differs from previous work by seeking times when the magnetic field angle is stable for a time duration. The reason for selecting for magnetic field is so that this study can distinguish between switchback patches and larger scale quiescent wind. Previous work selected quiescent wind by using the $z$ threshold of $z < 0.05$ and so captured quiescent wind between individual switchbacks as well as streams of quiescent wind between larger scale switchback patches.

To distinguish between streams quiescent wind and quiescent wind inside of switchback patches, it is necessary to bin $z$ in time. This binning process presents a challenge because the characteristic time lengths that should be used identify quiescent regions are unknown. At present, no study has quantified a characteristic quiescent region duration. Based on work on characteristic switchback timescales by \citet{deWit2020} and occurrence rates of switchback patches by \citet{Jagarlamudi2023ApJ} it is not clear that a typical duration does exist. 

To accurately identify quiescent regions in time, we employ an algorithm which places bins randomly in time across PSP encounters. In this algorithm and analysis, solar encounters are defined as times when PSP approaches within $65$ solar radii ($R_{\odot}$) of the sun. $z$ is calculated at the 4 Sa/cycle cadence, or roughly 4.53 Sa/s; the same downsampled cadence which we use for the magnetometer data, acknowledging that both the SPAN and SPC instruments record survey data at lower cadences than the magnetometer. As such, each data point from the magnetometer is assigned to its ``nearest neighbor" in the particle data to calculate $z$ for each point in the magnetometer data.

Once $z$ has been calculated for each encounter, 1000 points in $z$ are selected randomly to become the edges of bins in time, the gaps between these edges make up the width of the bin. The random selection helps eliminate systematic biases that may arise from bin sizing and bin placement. For each of these bins, the fraction of quiescent data points in each bin is determined. Quiescent data points are those which have $z$ values of $< 0.05$ or $> 0.95$. If this fraction of quiescent data within the bin is $> 0.95$ then each time step in the bin is assigned a 1, if the fraction is less, it receives a 0. A single iteration of this algorithm generates a line featuring square pulses with height 1. This procedure is repeated $N$ times and the average of each iteration $q_n$ (the first iteration being $q_0$) is taken as a final quality flag $q_{rand}$ over time $\tau$ which will be used to identify quiescent regions.

\begin{figure}[!htb]
    \centering
    \includegraphics[width=0.95\textwidth]{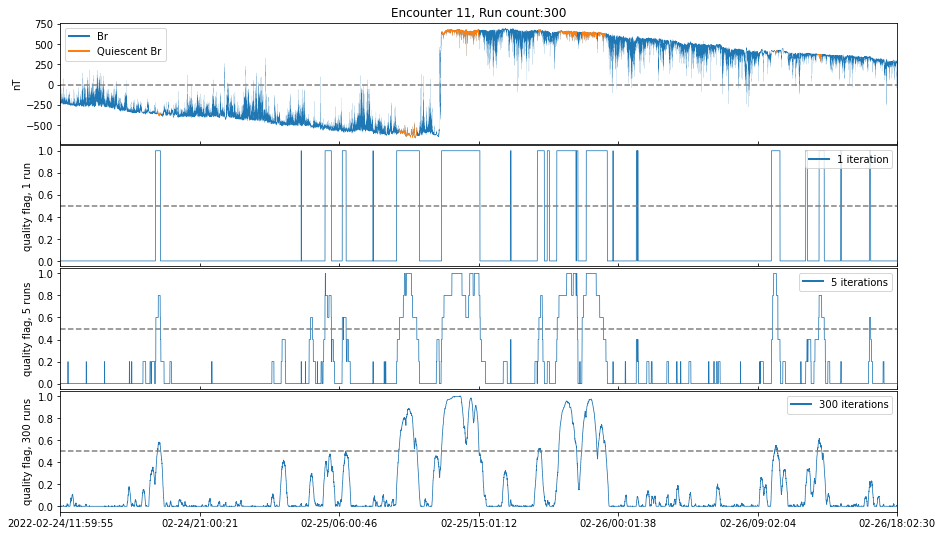}
    \caption{Quiescent region finding algorithm. Figure \ref{fig:algorithm}a is the magnetic field radial component in time series for encounter 7 with quiescent times highlighted in orange. Figure \ref{fig:algorithm}b is the result for quality flag $q_{rand}$ after one iteration of this identification algorithm. Figure \ref{fig:algorithm}c is the result after 5 iterations of the algorithm. Figure \ref{fig:algorithm}d is the result after 300 iterations. The final result for $q_{rand}$ used in this study is taken after 300 iterations. For panel a, the grey dashed line represents 0 nT. For panels b through d, the dashed grey line represents the 0.5 threshold required for solar wind to be labeled quiescent.}
    \label{fig:algorithm}
\end{figure}

Figure \ref{fig:algorithm} shows this algorithm acting on data from PSP encounter 11. Panel (a) displays the radial component of magnetometer survey data (blue) and quiescent regions identified by the final output of this algorithm (orange). Panels (b) though (d) show the quality flag output by this algorithm at increasing number of iterations. Panel (b) shows the result after one iteration, panel (c) after five iterations, and panel (d) shows the final output at $N = 300$ iterations.

\begin{displaymath}
    q_{rand}(\tau) = \displaystyle \frac{\sum_{n=0}^{N} q_n(\tau)}{N}
\end{displaymath}

\noindent Points in this quality flag with values $q_{rand} > 0.5$ are labeled quiescent solar wind and points with values $q_{rand} < 0.5$ are labeled non-quiescent. Using this metric, start and stop times are determined for each quiescent region. In panel (d) of Figure \ref{fig:algorithm}, we see several portions of solar wind which approach the $q_{rand} > 0.5$ threshold, but do not reach it. A set of examples are shown around 2022-02-25/06:00:00. This threshold of 0.5 is chosen in order to set it equally between the possible range of $q_{rand}$ between 1 and 0 and to better emulate the stepwise behavior of the initial $q_0$ iteration. This choice of threshold minimizes the risk of flagging false positives in quiescent identification, but throws away some portions of solar wind which may be on the edge of being quiescent.

The number of quiescent regions identified using this algorithm for encounters 1 through 16 are 504 and cover radial distances from 13.3 $R_{\odot}$ to 65 $R_{\odot}$. The total amount of time covered by these identified quiescent regions are nearly 494 hours of the total hours 5702 spent by PSP below 65 $R_\odot$. Quiescent regions thus make up $\sim 8.7\%$ of the total data available below 65 $R_\odot$. For this analysis, the data are broken into bins of radial distance. These radial bins are taken in 10 $R_{\odot}$ increments with the exception of the innermost bin. These bins are 13.3 $R_{\odot}$ to 15 $R_{\odot}$, 15 $R_{\odot}$ to 25 $R_{\odot}$, 25 $R_{\odot}$ to 35 $R_{\odot}$, and so on up to the 55$R_{\odot}$ to 65 $R_{\odot}$ radial bin. The statistics and data durations in hours are listed in Table \ref{tab:statistics}.

{}

Around 2022-02-25/12:00:00 is a Heliospheric current sheet (HCS) crossing followed by a large quiescent region highlighted in orange. We can see during this interval many switchbacks and switchback patches which are separated by quiescent regions. 


\subsection{SPAN-Ion Field-of-View Compensation and Quasi-Thermal Noise (QTN) Agreement} \label{subsec:FOV}

This investigation uses plasma moments derived from the Solar Wind Electrons Alphas and Protons (SWEAP) instrument suite \citep{Kasper2016}. Because Parker Solar Probe orbits so close to the Sun, it has a Thermal Protection Shield (TPS), or heat shield, to protect the spacecraft bus and its instruments from the intense light of the Sun \citep{Fox2016}. This heat shield blocks light from the Sun, but also partially blocks incoming solar wind particles which means we only measure partial plasma moments of solar wind temperature, density, and velocity. SWEAP accounts for this by using two different particle detectors in conjunction. On the forward end of the spacecraft sits the Solar Probe Cup (SPC), which is a sun-pointing Faraday cup that is limited to measuring particles in the radial direction, producing a reduced 1D velocity distribution function (VDF) \citep{Case2020}. The Solar Probe ANalyzer for ions (SPAN-I) sits on the ram of the spacecraft and can measure a full 3-dimensional proton VDF. However, because of the instrument positioning behind the heat shield, the core of the proton distribution is not always within the SPAN-I field-of-view, leading to inaccuracies in determining solar wind bulk properties under some circumstances \citep{Livi2022ApJ}. The proton core drifts out of the SPANs field-of-view when the vector of the proton flow velocity in the PSP frame is oriented more along the radial direction than the tangential direction, and occurs more often when PSP is moving slowest in the tangential direction relative to the Sun-PSP line. 

In this study we investigate thermal properties which require the proton VDF in 3-dimensions; namely, the solar wind temperature anisotropy and the plasma beta parallel to the background magnetic field. To ensure that our results are robust to field-of-view (FOV) changes in the SPAN-I data, we control for the field-of-view in two ways: the first is by observing the energy partition in SPANs anodes along its Sunward-Antisunward viewing directions. When the peak of the energy flux is within the 3rd anode an accurate fit of the core is possible \citep{Livi2022ApJ}. Thus times when the peak energy flux is within SPANs 3rd anode from the heat shield are considered accurate and labeled with a 2, while times where the peak flux is not within the 3rd anode are labeled with a 0. For the purposes of ease of data analysis, these times are averaged across using a 30 minute long sliding window. The 30 minute window is chosen to select times when PSP enters large scale solar wind streams in which the proton core is within SPANs FOV. This gives us a partial quality flag which we call $q_{fov}$ which varies in a binary fashion between 2 and 0.

The second way in which SPAN data quality is controlled for is by comparing the estimated proton density measured by SPAN (called $n_p$) to the value for the electron density ($n_e$) inferred from quasi-thermal noise (QTN) measurements of the upper hybrid resonance line. By comparing the difference between the $n_e$ and $n_p$ and assuming quasi-neutrality, we get an estimation on how accurately SPAN-I is measuring the proton distribution. An accurate measurement of the proton density $n_p$ by SPAN-I is a strong indication that SPAN-I is getting a accurate fit of the proton core and thus, at those times, an accurate measurement of the solar wind velocity vector. This comparison gives For each point in the QTN measurement there is an uncertainty in the density which originates from the frequency resolution of the FIELDS instrument suite. This uncertainty is $\delta n_e$. For times when the QTN estimates are available, each point in the QTN data is assigned to its nearest neighbor in the SPAN data using the same method described in Section \ref{subsec:algorithm}. The percent error $\delta n_e/n_e*100$ on the QTN density $n_e$ has typical values of $\sim 8\%$ and thus $2 \times \delta n_e$ is around $\sim 16\%$ of the measured value. A SPAN density $n_p$ within $2 \times \delta n_e$ is considered most accurate and assigned a 2. If $n_p$ is within $3 \times \delta n_e$ of $n_e$ it is considered somewhat accurate and assigned a 1. Anything else, whether $n_p$ is outside of $3 \times \delta n_e$ or if the QTN data is unavailable, that time is considered inaccurate and is assigned a 0. This gives us a partial quality flag which we call $q_{qtn}$ which varies between 2, 1, and 0. Note that this QTN/SPAN density comparison does not consider alpha abundance. So long as the alpha abundance remains below $~4 \%$ the alphas should not have an exceptionally large effect on this analysis given the comparatively large window $2 \times \delta n_e$, which is typically around $\sim 16\%$ of $n_e$, allowed for this comparison. However, it does mean that times with high alpha abundance, if the proton core is not sufficiently within the field of view as determined by $q_{fov}$, will be cut out of the analysis and may serve to bias this analysis to low alpha solar wind.

These two methods described above are combined into a total quality flag for SPAN. The values for $q_{fov}$ and $q_{qtn}$ are added to create a total quality flag for SPAN which we call $q_{SPAN} = q_{fov} + q_{qtn}$. This total quality flag varies between integer values 4, 3, 2, 1, and 0. To calculate the Parker Spiral unit vector for $z$, SPAN-I is used whenever this quality flag $q_{SPAN}$ is $\ge1$ and SPC is used otherwise. This occurs when the SPAN density is within $3 \times \delta n_e$ or when the peak of the SPAN distribution is within the 3rd anode in the Sunward-Antisunward viewing directions or some combination of both situations. For all other measurements which rely on SPAN in this study such as the bulk velocity, proton temperature, proton density, proton temperature anisotropy $\mathcal{R}$, proton plasma beta parallel $\beta_{\parallel p}$, and Alfvén mach number, a stricter requirement of $q_{SPAN} \ge 2$ is required. This stricter requirement ensures that either the proton core is well within the SPAN-I FOV or that the SPAN proton density $n_p$ is within $2 \times \delta n_e/n_e *100\%$ of $n_e$ or some combination of both situations.

\begin{table}
    \begin{tabular}{|c|c|c|c|c|c|c|}
         \hline
         $\#$ of Regions (Hours) & $<15$ $R_{\odot}$ & 15-25 $R_{\odot}$ & 25-35 $R_{\odot}$  & 35-45 $R_{\odot}$ & 45-55 $R_{\odot}$ & 55-65 $R_{\odot}$ \\ [0.5ex]
         \hline\hline
         Raw Region Count & 44 (35.6)  & 92 (109.3) & 107 (114.3) & 128 (107) & 67 (79.4) & 62 (48.7) \\
         \hline
         After SPAN Check & 44 (31.4) & 87 (72.7) & 96 (49.3) & 49 (15.7) & 30 (12)  & 6 (1.1) \\
         \hline
    \end{tabular}
    \caption{Table of quiescent region counts and hours of data available for different PSP radial bins. The number of regions are integers to the left while the number of hours are contained in parenthesis on the right of each cell. Top row is the counts of regions found at all radial distances in the magnetic field data. Bottom row are the number of regions for which the SPAN-I field of view is sufficient.}
    \label{tab:statistics}
\end{table}

\subsection{Magnetic Field Fluctuation Amplitudes: $\delta B/\langle|B|\rangle$} \label{subsec:delb}

From previous studies, quiescent regions possess different turbulent properties than switchbacks \citep{deWit2020} and the quiescent regions have lower amplitude fluctuations but this property has not been fully quantified to date. We can estimate the amplitude of the turbulent fluctuations by employing a $\delta B / \langle |B| \rangle$ estimation. $\delta B / \langle |B| \rangle$ is the ratio of the average amplitude of magnetic fluctuations relative to the average background magnetic field. $\delta B$ is defined as

\begin{displaymath}
    \delta B = \sqrt{\langle \delta {B_r}^2 + \delta {B_t}^2 + \delta {B_n}^2 \rangle}
\end{displaymath}

\noindent where the fluctuations of the i-th component are defined to be

\begin{displaymath}
    \delta B_i = B_i - \langle B_i \rangle
\end{displaymath}

\noindent For quiescent region, the above operation is used to return a single value. Because it involves many averages, we must choose a scale over which to evaluate that average. The scale by which we determine the average is the scale of the quiescent region itself. For example, if a particular quiescent region is 10 minutes long, the scale over which the average is determined is 10 minutes. Likewise, if the next region is 2 hours long, the scale of the average for that region is 2 hours. For each quiescent region a value of $\delta B / \langle |B| \rangle$ is determined. Then using the same set of averaging scales, portions of the remaining solar wind not identified as quiescent regions are selected at random and a value for $\delta B / \langle |B| \rangle$ is calculated for non-quiescent wind in the same way as the quiescent wind. This is done in order to compare quiescent and non-quiescent magnetic field fluctuation amplitudes in a way that minimizes bias in bin placement. For example, if we calculated $\delta B / \langle |B| \rangle$ for a quiescent region and, keeping the same averaging scale, calculated $\delta B / \langle |B| \rangle$ for non-quiescent wind immediately following in time series, it may be the case that the value of $\delta B / \langle |B| \rangle$ for both sets of solar wind are correlated by virtue of proximity. Randomly selecting where to place the non-quiescent wind $\delta B / \langle |B| \rangle$ calculations helps capture the general trend of $\delta B / \langle |B| \rangle$ in non-quiescent wind more accurately. To increase the amount of statistics in the non-quiescent distribution, this calculation is done as many times as non-quiescent wind over an encounter is able to be divided. This allows for direct comparisons of quiescent region turbulence amplitudes with non-quiescent solar wind.

\section{Results} \label{sec:results}

\subsection{Velocity Differences} \label{subsec:vel}

\begin{figure}[!htb]
    \centering
    \includegraphics[width=0.95\textwidth]{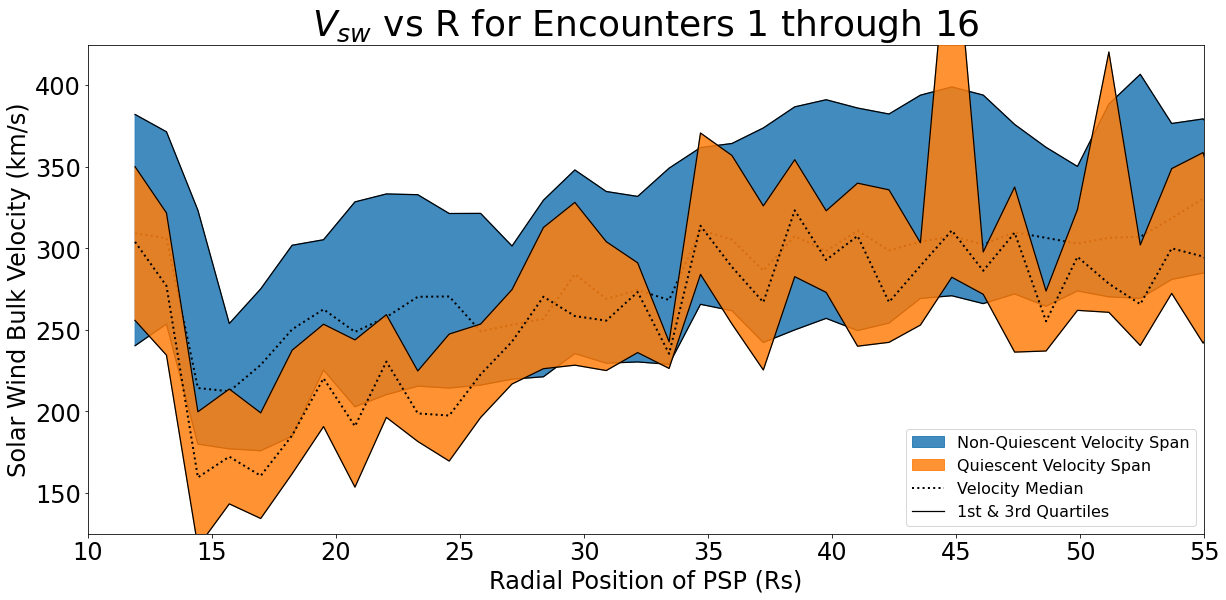}
    \caption{Bulk Flow Velocity span across entire PSP mission inside of 55 Rs vs radial distance. Each curve represents a 2d-histogram of the solar wind velocity measured by either SPAN or SPC depending on quality of data as described in Section \ref{subsec:FOV}. The top and bottom solid black lines bounding the colored areas represent the 1st and 3rd quartiles in each radial bin, while the dotted line in the center of each colored area is the median value for the solar wind speed. Quiescent regions data is highlighted in orange where non-quiescent wind is highlighted in blue. The velocity spread for quiescent regions are consistently lower for most altitudes.}
    \label{fig:velocity}
\end{figure}

\begin{figure}[!htbp]
    \centering
    \includegraphics[width=0.95\textwidth]{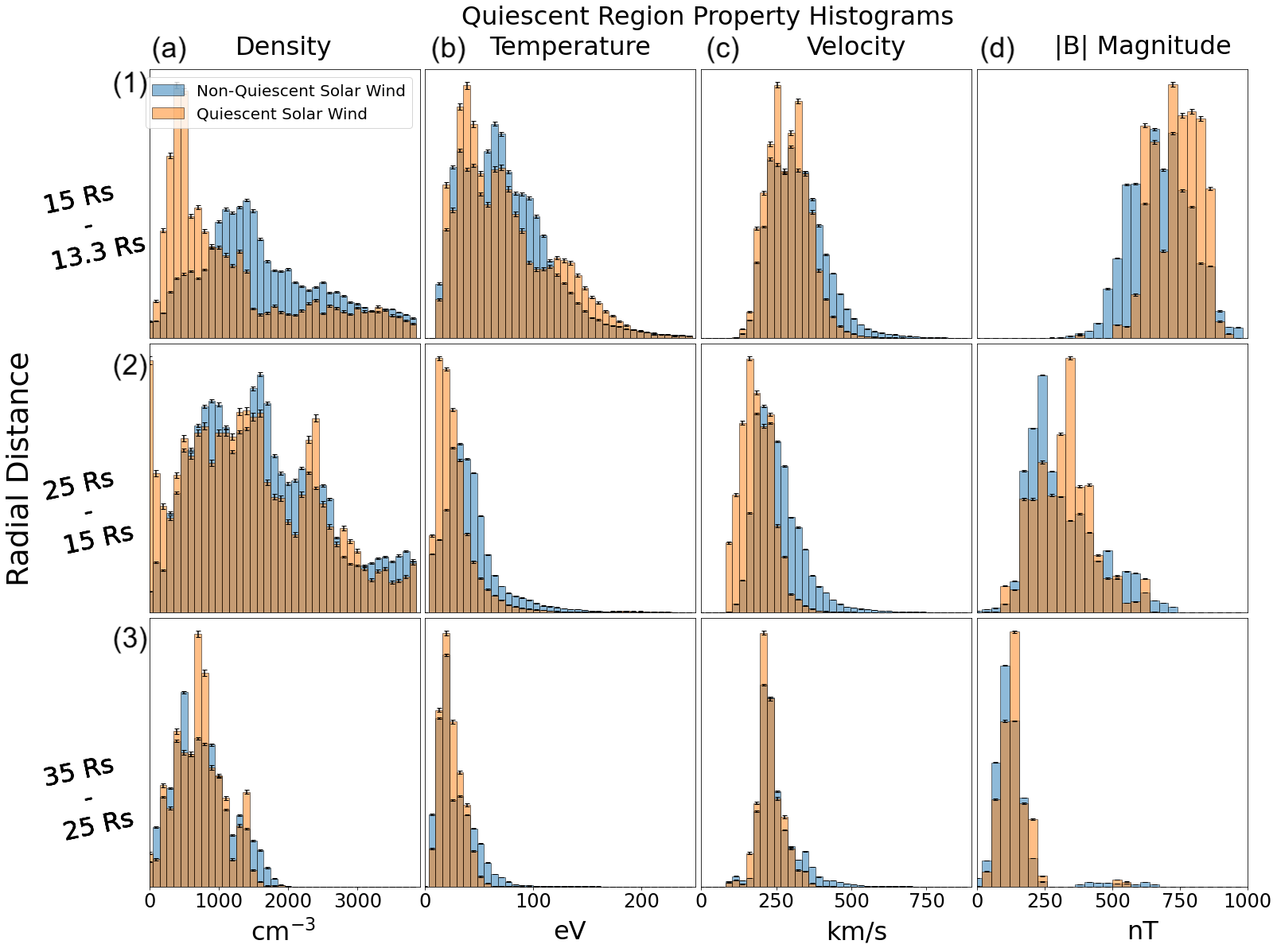}
    \caption{Normalized histograms of various solar wind properties. The rows of histograms represent radial distances with the top row being the innermost radial bin with data ranging from 13.3 $R_{\odot}$ to 15 $R_{\odot}$. The following two rows are in 10 $R_{\odot}$ increments, the center row being from 15 to 25 $R_{\odot}$ and the bottom row from 25 to 35 $R_{\odot}$. The columns are each property at those radial distances. Column (a) is the number of protons per cm$^{3}$, column (b) is the proton core temperature in eV, column (c) is the bulk velocity magnitude in km/s, column (d) is the magnetic field magnitude in nT.}
    \label{fig:histograms}
\end{figure}

Figure \ref{fig:velocity} shows solar wind bulk flow velocity data from the entirety of the Parker Solar Probe mission up to encounter 16 plotted against PSPs heliocentric distance from the sun. Figure \ref{fig:velocity} represents 2d histograms for quiescent and non-quiescent wind over the course of PSPs mission. Quiescent region data are pictured in orange and non-quiescent wind is depicted in blue. The solid lines at the top and bottom of each shaded area represent the 1st and 3rd quartiles for each distribution and the dotted line in the center of each shaded curve is the median.  From this figure, we can see that quiescent times make up the slower end of the solar wind velocity while non-quiescent times tend towards higher velocities. This observation becomes more clear when comparing to the third column of Figure \ref{fig:histograms}. Figure \ref{fig:histograms} is four sets of three histograms for proton density, proton core temperature, bulk flow velocity, and the magnetic field magnitude in order from left column to right column. The rows of Figure \ref{fig:histograms} represent three radial bins which are $<15$ $R_\odot$, $15$ $R_\odot$ to $25$ $R_\odot$, and $25$ $R_\odot$ to $35$ $R_\odot$ from top row to bottom row. This figure contains data for the innermost three radial bins as outside of 35 $R_{\odot}$, the amount of SPAN data which satisfies the quality flag in Section \ref{subsec:FOV} drops off quickly. 

At each radial distance, the average solar wind velocity for quiescent regions is lower than the average of non-quiescent times. For Figure \ref{fig:histograms}c1 the average quiescent speed is 302 km/s and the average speed during non-quiescent times is 333 km/s, which is a 31 km/s difference or $\sim 10 \%$ change. For Figure \ref{fig:histograms}c1, it is possible that the observed long tail for non-quiescent regions may be a spacecraft observation effect: higher fluctuations in the perpendicular directions can create larger uncertainties due to sampling geometry. This change is outside the uncertainty for reported velocities from SPAN-I moments which is $\sim 3 \%$, assuming good SPAN-I FOV. However, this uncertainty estimate does not include systematic uncertainties not yet quantified by the instrument. At larger radial distances the averages begin to diverge. When looking at the 15 to 25 $R_{\odot}$ radial bin in Figure \ref{fig:histograms}c2 the average quiescent speed becomes 198 km/s and the average speed during non-quiescent times becomes 266 km/s. The difference in solar wind speeds expands to a 68 km/s difference, or a change of $\sim 34 \%$ from the quiescent solar wind speed. This change in the percent differences could indicate differential acceleration between quiescent and non-quiescent wind populations across the observed radial bins. However, this possibility warrants further investigation because while the differences in solar wind speeds between the two populations increase between radial distances the innermost bin also appears to have higher solar wind speeds overall. This could be a result of selection biasing due to the innermost bin being dominated by data at solar maximum. Possible biasing due to changes in the solar cycle is discussed in Section \ref{subsec:evolution}.

{}

{}

\subsection{Temperature Differences} \label{subsec:temp}

In addition to the differences in velocity, there is also a temperature difference between quiescent and non-quiescent solar wind. In Figure \ref{fig:histograms}b2, quiescent regions make up the lower end of the temperature range. However, for the other two radial bins, Figure \ref{fig:histograms}b1 and Figure \ref{fig:histograms}b3, the difference becomes more ambiguous. For Figure \ref{fig:histograms}b1, the average for both quiescent and non-quiescent temperatures are 79.3 eVs. Then stepping out in radial distance to Figure \ref{fig:histograms}b2, a difference in temperatures emerges, with quiescent times averaging 31.6 eV and non-quiescent solar wind averaging 47.2 eV. This represents a $\sim 49 \%$ increase in temperature, which is outside the SPAN moments uncertainty in temperature which is $\sim 10 \%$ to $\sim 15 \%$, though for non-quiescent solar wind this uncertainty may be higher due to increased magnetic field fluctuations \citep{Verscharen2011}. Then finally Figure \ref{fig:histograms}b3, quiescent regions average 26.9 eV whereas non-quiescent wind averages 30.2 eV, which yields a $\sim 12 \%$ increase in temperature. This increase is within the SPAN-I uncertainty values for temperature. A difficulty with these results is that the temperature in the lowest altitude radial bin is multi-modal. This may indicate too little quiescent region data (44 quiescent regions, 35.6 hours of data) in the range of altitudes that are being investigated. While the number of individual data points per quiescent region are high, the number of quiescent regions within this altitude range may not be enough to capture the full nature of the temperature distribution for quiescent regions at this altitude range. More orbits within 15 $R_{\odot}$ are needed to fill in this distribution.

\begin{table}[!htb]
    \centering
    \begin{tabular}{|c||c|c|c|}
        \hline
         Q-region , Non-Q & 13.3 $R_{\odot}$ - 15 $R_{\odot}$ & 15 $R_{\odot}$ - 25 $R_{\odot}$ & 25 $R_{\odot}$ - 35 $R_{\odot}$\\ [0.5ex] 
        \hline\hline
        Density (cm$^{-3}$) & 1288 ± 1057, 2207 ± 1732 & 1904 ± 1434 , 2029 ± 1421 & 776 ± 354, 800 ± 407\\
        \hline
        Temperature (eV) & 79.3 ± 46.9 , 79.3 ± 57.2 & 31.6 ± 23.5, 47.2 ± 57.4 & 26.9 ± 10.3 , 30.2 ± 16.6\\
        \hline
        Velocity (km/s) & 302 ± 70 , 333 ± 97 & 198 ± 54 , 266 ± 82& 240 ± 42 , 259 ± 67\\
        \hline
        $|B|$ (nT) & 752.6 ± 84.9 , 701.9 ± 128.5 & 339.2 ± 107.8 , 335.8 ± 135.4 & 154.7 ± 65.8 , 149.0 ± 89.6\\
        \hline
    \end{tabular}
    \caption{Solar wind quantity averages for Figure \ref{fig:histograms} with standard deviations included. Quiescent solar wind averages are on the left side of the comma in each cell, non-quiescent solar wind is on the right. The rows represent the solar wind properties, the columns represent the radial distance.}
    \label{tab:averages}

\end{table}

\begin{figure}[!htb]
    \centering
    \includegraphics[width=0.75\textwidth]{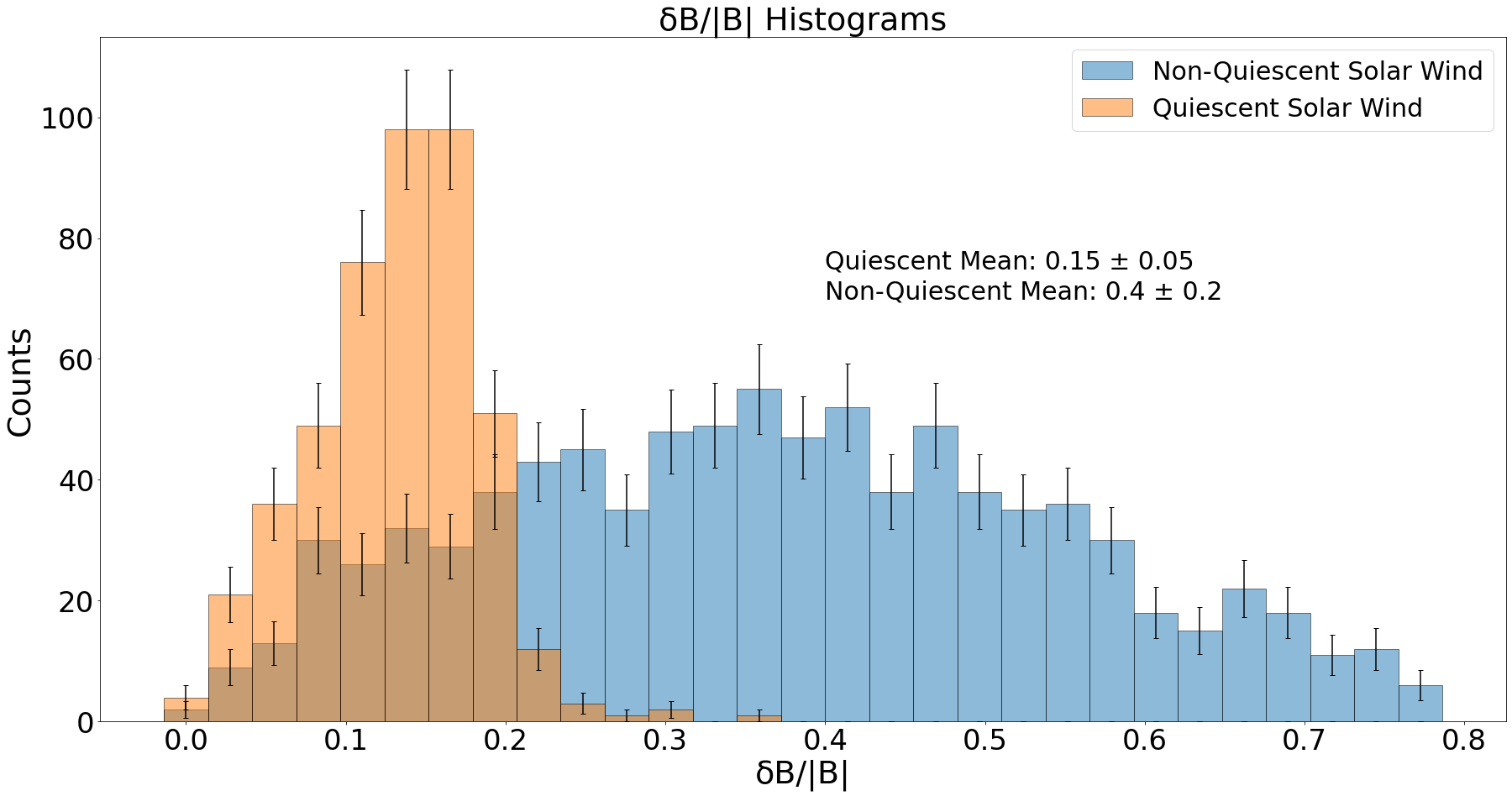}
    \caption{Non-normalized histograms of $\delta B/\langle|B|\rangle$ for each quiescent region across all radial distances. The orange histogram is for quiescent regions and the blue is for non-quiescent solar wind of the same scale.}
    \label{fig:delb}
\end{figure}{}

\subsection{Magnetic Field Differences} \label{subsec:magnetic}

From Figure \ref{fig:histograms}d1, the DC magnetic field magnitude is on average greater for quiescent regions than non-quiescent wind at similar altitudes. The averages are enumerated in Table \ref{tab:averages}. From both Table \ref{tab:averages} and from each panel in Figure \ref{fig:histograms}d for each radial distance the magnetic field is greater on average than for non-quiescent intervals. Moreover, from Figure \ref{fig:delb} it is shown that the turbulence amplitudes for quiescent regions relative to the background field are lower than for non-quiescent solar wind. Figure \ref{fig:delb} are a set of two histograms, one for quiescent regions in orange, one for non-quiescent wind in blue. The histograms represent the magnetic field fluctuation amplitudes $\delta B / \langle |B| \rangle$ for quiescent and non-quiescent wind. The average magnetic field fluctuation amplitudes $\delta B / \langle |B| \rangle$ for quiescent regions are 0.15±0.05 whereas the value is 0.4±0.2 for non-quiescent solar wind. This amounts to a $62.5 \%$ decrease in $\delta B / \langle |B| \rangle$ in quiescent regions compared to non-quiescent wind in this study.

\subsection{Density Differences} \label{subsec:dens}

With the exception of the innermost bin, the density measured is roughly comparable. From Figure \ref{fig:histograms}a2 and \ref{fig:histograms}a3, it is shown that the distributions for the density measurements, while not identical, are similar in form and in position between the quiescent and non-quiescent wind. However, when looking at the 13.3 $R_{\odot}$ - 15 $R_{\odot}$ radial bin, the distribution for quiescent times peaks in an entirely different place when compared to non-quiescent solar wind and its averages are lower. This bin shows average densities of 1288 cm$^{-3}$ for quiescent regions and 2207 cm$^{-3}$ for non-quiescent wind. However, like the temperature measurement, this difference may be a result of low quiescent region counts in that particular radial bin. Further orbits will help clarify the validity of this observation.

\begin{figure}[ht!]
    \centering
    \begin{minipage}{0.48\textwidth}
        \centering
        \includegraphics[width=1\textwidth]{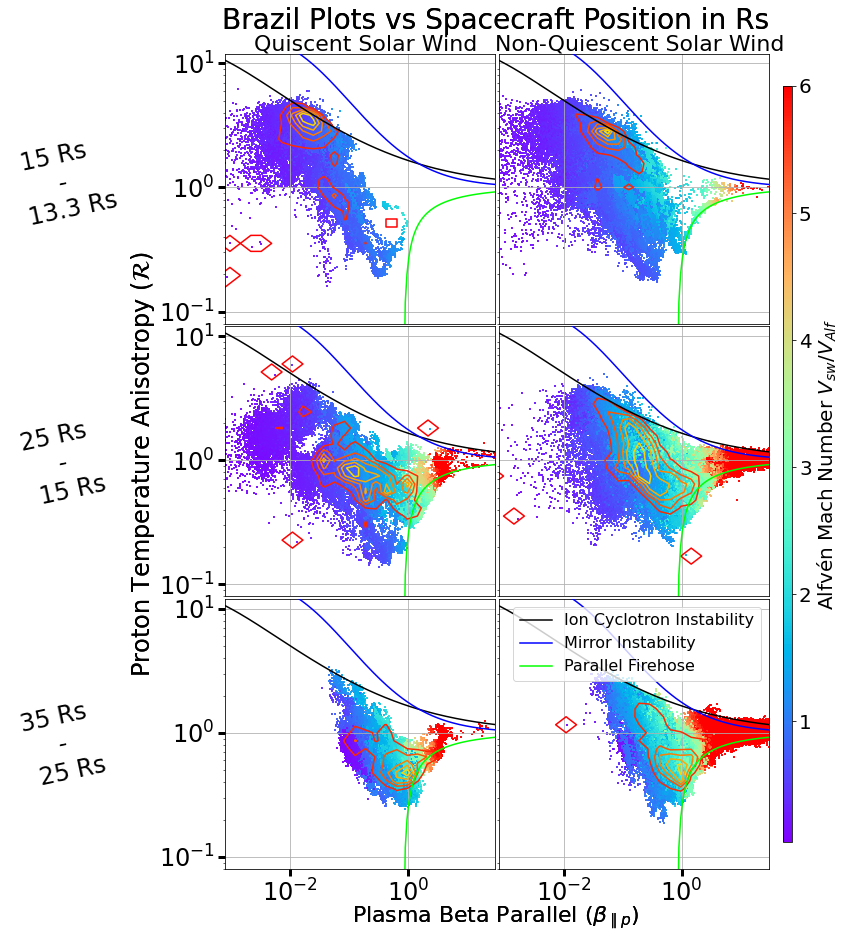} 
        \caption{Scatter plots of proton $\mathcal{R}$ vs $\beta_{\parallel p}$ for quiescent and non-quiescent solar wind at different radial distances. The y-axes of the subplots are the proton temperature anisotropies and the x-axes of the subplots are the plasma betas parallel to the background magnetic field. The columns of this plot represents distributions for quiescent plots on the left and non-quiescent plots on the rightmost column. The rows represent radial bins with the topmost row being the innermost altitudes that PSP reaches and lower rows represent higher altitudes. The color scale is the Alfvén mach number and the contours represent point density.}
        \label{fig:brazil}
        
    \end{minipage}\hfill
    \begin{minipage}{0.5\textwidth}
        \centering
        \includegraphics[width=1\textwidth]{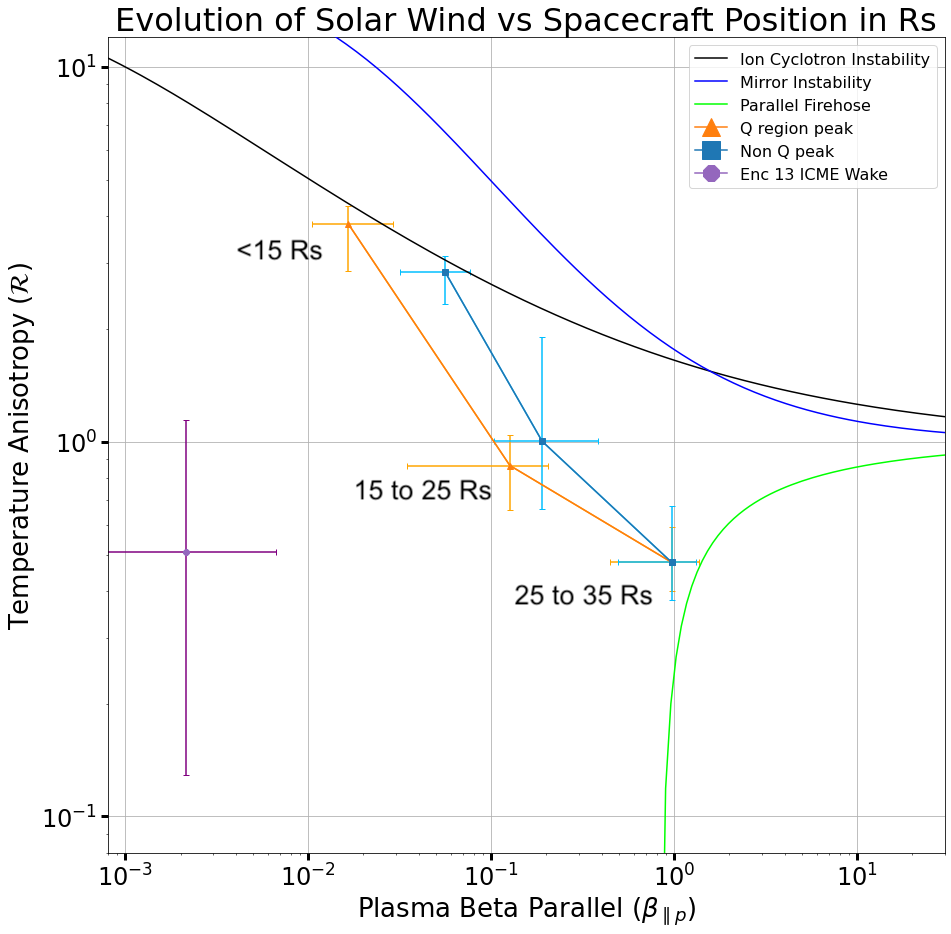} 
        \caption{Ion $\mathcal{R}$ vs $\beta_{\parallel p}$ plot for Figure \ref{fig:brazil} peaks. The y-axis is the proton temperature anisotropy and the x-axis is the plasma beta parallel to the background magnetic field. The peaks of each subplot in Figure \ref{fig:brazil} are tracked through altitudes and plot relative to one another. Theencounter 13 ICME wake is plotted in purple, this event is discussed in Section \ref{subsec:evolution}. Solid lines are various instability thresholds past which solar wind would become likely to grow waves. The black line is the ion cyclotron instability, the blue is the mirror mode instability, and the green line is the parallel firehose instability \citep{Hellinger2006}. These peaks are based on 2d histograms of the data and the error bars are full width of the distributions at half maximum.} 
        \label{fig:brazilpeaks}
    \end{minipage}
\end{figure}

\subsection{Ion Temperature Anisotropy ($\mathcal{R}$) vs Plasma Beta ($\beta_{\parallel p}$) and Alfvénic Instabilities} \label{subsec:brazil}

This study also investigates proton plasma beta $\beta_{||p}$ and proton temperature anisotropies $\mathcal{R}$. Figures \ref{fig:brazil} and \ref{fig:brazilpeaks} are log-scale scatter plots of solar wind in $\mathcal{R}$ vs $\beta_{\parallel p}$ parameter space, where $\mathcal{R} \equiv T_{\perp p}/T_{\parallel p}$ and $\beta_{\parallel p} \equiv 2 \mu_0 n k T_{\parallel p} / B^2 $, where the proton properties $n, T_{\parallel p}$ are taken from proton measurements of SPAN-I. These plots quantify the evolution of quiescent regions and non-quiescent wind and relate these solar wind populations to plasma instabilities. Figures \ref{fig:brazil} and \ref{fig:brazilpeaks} represent different aspects of the same data. The left column of Figure \ref{fig:brazil} is a set of $\mathcal{R}$ vs $\beta_{\parallel p}$ plots including every point of data for which the proton core peak is within the SPAN FOV. These scattered points are colored by the Alfvén mach number, the ratio of the solar wind velocity magnitude in the Heliocentric RTN frame to the local Alfvén speed. The contours overplotted onto the scatter plots represent normalized point distributions where the contours bound regions in the parameter space where points are most concentrated. Figure \ref{fig:brazilpeaks} on the right displays the positions of the peaks of the $\mathcal{R}$ vs $\beta_{\parallel p}$ distributions as they are measured at varying distances from the sun and tracks them through the parameter space. From both plots, it is observed that both the quiescent and non-quiescent solar wind are well-bounded by the ion cyclotron instability and the parallel firehose instability thresholds \citep{Hellinger2006}. From Figure \ref{fig:brazilpeaks}, it is observed that the peaks of the quiescent regions distributions are consistently farther into the stable portion of parameter space, to the left of the non-quiescent solar wind. From Figure \ref{fig:brazil}, it is observed that quiescent region solar wind has a lower Alfvén Mach number than non-quiescent solar wind at similar portions of the $\mathcal{R}$ vs $\beta_{\parallel p}$ parameter space. In both Figures \ref{fig:brazil} and \ref{fig:brazilpeaks} for the lowest altitude bin, quiescent regions start from a different portion of parameter space but evolve along a similar trajectory as non-quiescent wind. The distribution peaks in Figure \ref{fig:brazilpeaks} are generated using a 2d-histogram of the scatter plot data in Figure \ref{fig:brazil}. Because of the use of histograms, determined peak position can be affected by where bins are placed and the size of those bins. The 2d histograms used to generate this plot are NxN grids in log space and the number N chosen in Figures \ref{fig:brazil} and \ref{fig:brazilpeaks} is 32. In order to ensure that the trends observed in Figure \ref{fig:brazilpeaks} are not a result of binning, this plot has been generated using varied bin counts from N=10 to N=38. When changing the bin numbers the exact peak positions and error bars do shift, but the general results shown in Figure \ref{fig:brazilpeaks} remain the same: quiescent wind begins at a different position in the $\mathcal{R}$ vs $\beta_{\parallel p}$ space, with greater $\mathcal{R}$ and lower $\beta_{\parallel p}$ and evolve along similar trajectories while remaining in the more stable part of the space.

\vspace{-1cm}

\section{Discussion} \label{sec:discussion}



\subsection{Quiescent Region Turbulence and the Possibility of In-Situ Origins.} \label{subsec:in-situ}

Figure \ref{fig:delb} demonstrates that quiescent regions have magnetic field fluctuation amplitudes $\delta B / \langle |B| \rangle$ lower than non-quiescent solar wind. Along with lower amplitude field fluctuations, quiescent regions have lower bulk flow velocity and cooler proton temperatures in the radial bins which have the most statistics. This cooler temperature has implications for the possible origins of quiescent regions. We suggest two distinct possibilities exist for quiescent region origins: either they are generated in-situ as plasma travels from the photosphere into the corona consistent with hypothesis 3 in Section \ref{subsec:introorigins}, or they are launched from very close to the solar surface and better retain their turbulent and bulk properties as they travel to the spacecraft consistent with hypotheses 1 and 2 in Section \ref{subsec:introorigins}.

If quiescent regions developed on the way to the spacecraft, this suggests the existence of a baseline level of magnetic turbulence which, in quiescent regions, is either damped, or is otherwise evacuated from the quiescent region. In the case of turbulent damping, suppose there exists a property of the solar wind in quiescent regions that causes Alfvénic fluctuations to be damped, the damping of these fluctuations would presumably place some or all of their wave energy into heating the protons in the quiescent regions leading to an increase in proton temperature relative to non-quiescent wind or would contribute to acceleration of the wind leading to an increase in velocity relative to non-quiescent wind. However, from Figure \ref{fig:histograms} it is observed that the opposite is the case and quiescent regions maintain a similar or cooler temperature than non-quiescent wind and are slower than more turbulent populations. Therefore we rule out this method of in-situ generation.

\subsection{Region Evolution and Plasma Instabilities} \label{subsec:evolution}

From Figures \ref{fig:brazil} and \ref{fig:brazilpeaks}, it is shown that quiescent regions occupy a more stable portion of the $\mathcal{R}$ vs $\beta_{\parallel p}$ parameter space. With decreasing radial distance, quiescent solar wind diverges from non-quiescent wind in the parameter space. This may suggest that closer to the source of quiescent regions the solar wind conditions are different. If these results can be extrapolated further down into the solar corona, it may suggest that quiescent regions emerge from field lines associated with low $\beta_{\parallel p}$ and high $\mathcal{R}$. Despite their different turbulent properties, both populations appear to evolve through the $\mathcal{R}$ vs $\beta_{\parallel p}$ parameter space along similar trajectories, suggesting that the same process dominates the evolution of both. Figures \ref{fig:brazil} and \ref{fig:brazilpeaks} show that the trajectories of the quiescent and non-quiescent plasma through the parameter space approach one another as Parker Solar Probe moves radially outward, which could imply a second evolution process driving the two populations together or that the two populations are affected differently by the primary process. 

However, these changes could also be a result of identifying different populations of quiescent-like wind that occur at different radial distances. For instance, during PSPs 13th orbit, a large Interplanetary Coronal Mass Ejection (ICME) passed over PSP. In the following days, PSP observed a region where the plasma density dropped significantly in the wake behind the path of the ICME \citep{Romeo2023ApJ}. This region of low plasma density in the wake of the ICME was marked as quiescent by the detection algorithm described in Section \ref{subsec:algorithm}. Because this plasma density dropout was clearly associated with the ICME and its position in the $\mathcal{R}$ vs $\beta_{\parallel p}$ parameter space is substantially different from the rest of the quiescent regions (shown in Figure \ref{fig:brazilpeaks}), it was excluded from the dataset used in this study. If multiple sources for quiescent wind exist; for example if some quiescent wind emerges from between switchback patches as in \cite{Bale2019,Bale2021,Bale2023} and others emerge from sources like the encounter 13 ICME, the properties of those sources would be mixed in this study.

The apparent convergence of the trajectories of quiescent and non-quiescent wind with radial distance seen in Figures \ref{fig:brazil} and \ref{fig:brazilpeaks} may also be an effect of selection biasing with respect to the solar cycle. The innermost radial bin is dominated by later solar encounters 10 through 16 which are more towards solar maximum than earlier solar encounters 1 through 9. PSP flies through varied solar wind streams as the Sun approaches solar maximum. PSP sampling these different streams may be causing a difference in the properties of solar wind which we observe as an increasing difference in the $\mathcal{R}$ vs $\beta_{\parallel p}$ positions for quiescent and non-quiescent wind in the innermost radial bin.

\begin{figure}[!htb]
    \centering
    \includegraphics[width=0.75\textwidth]{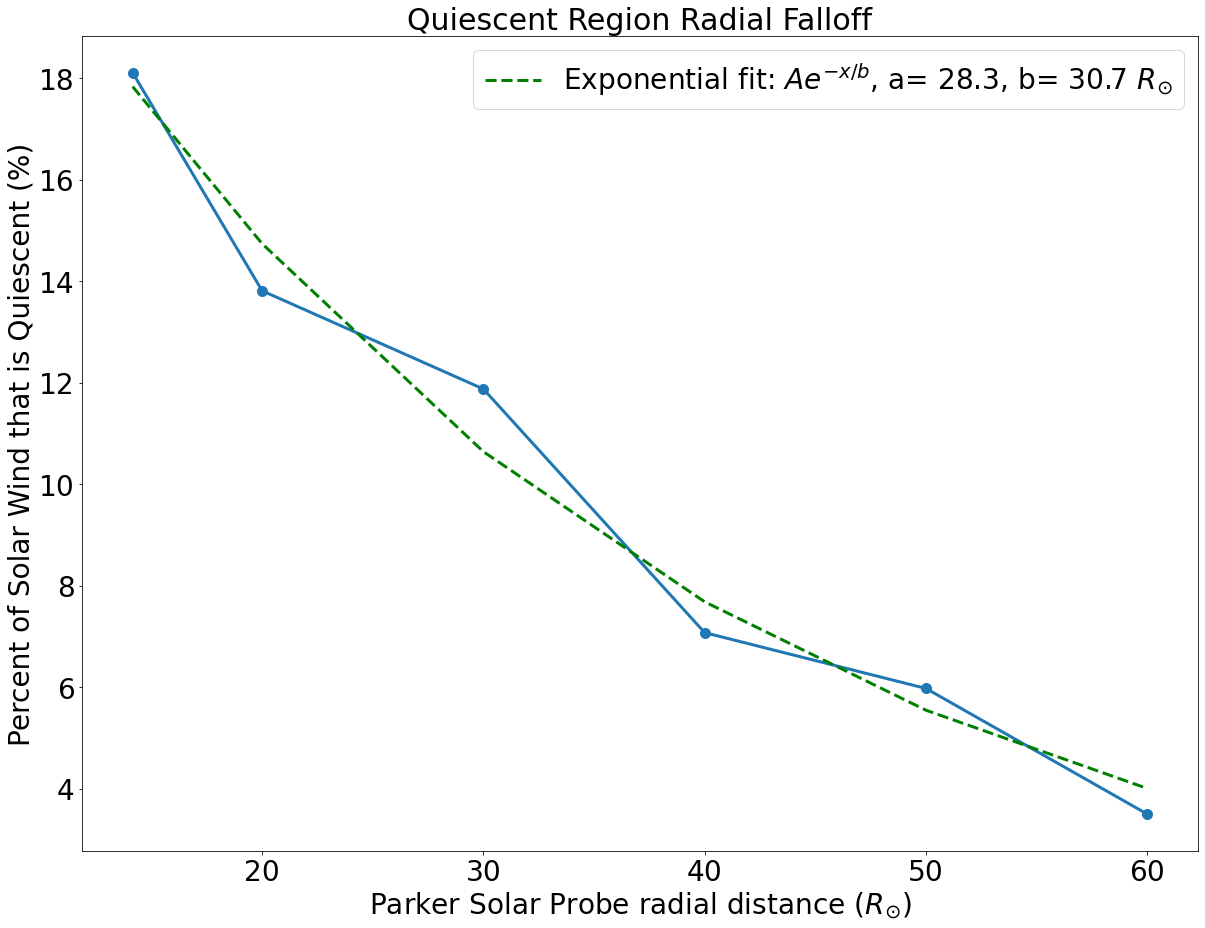}
    \caption{Radial evolution of quiescent region occurence. The vertical axis represent the percentage of solar wind plasma which has been labeled as quiescent out of the total plasma that PSP has flown through. The percentages presented are ratios of the quiescent meters of plasma flown through over the total meters of plasma flown through at that altitude. The fit exponential falls off with a scale of $\sim 30.7$ $R_{\odot}$ indicating a rapid decay of quiescent plasma with solar wind distance.}
    \label{fig:falloff}
\end{figure}{}

Quiescent regions are not reported in studies from missions which orbit farther out in the solar system. In order to pull out the shape of the Parker Spiral in measurements taken at 1 AU, one must average over some timescale \citep{Ness1964}. However, with these regions, their fields are roughly along the Parker Spiral direction without the need to average. The fact that missions farther out do not observe quiescent regions suggests that quiescent regions must decay, either by internally developing magnetic field fluctuations or else merging with the surrounding ambient solar wind turbulence. We attempt to quantify this decay rate by observing the meters of plasma flown through by PSP as a function of radial distance. Figure \ref{fig:falloff} shows the percentage of meters of plasma that quiescent regions make up at each radial bin described in Table \ref{tab:statistics}. The points in Figure \ref{fig:falloff} are placed at the center of each radial bin. An exponential is fit to the curve to roughly quantify a ``scale height" for quiescent regions. This fit reveals an exponential with a $30.7$ $R_\odot$ distance scale. This rapid decrease in quiescent plasma helps explain why these regions have not been reported on during previous missions. If the exponential fit in Figure \ref{fig:falloff} is extrapolated out to $60$ and $65$ $R_{\odot}$, the altitudes of Solar Orbiter and the Helios probes, the fit predicts $\sim 3.4 \%$ of the solar wind should be quiescent. With other missions like Bepi-Columbo orbiting at Mercury ($\sim86$ $R_{\odot}$), missions orbiting at Venus ($\sim155$ $R_{\odot}$), and many at $\ge 1$ AU ($\sim 215$ $R_{\odot}$), it is unsurprising that these regions have gone unreported by previous studies.

However, the mechanism by which quiescent regions decay remains unknown. One possibility, based on the trajectories shown Figure \ref{fig:brazilpeaks}, is that quiescent regions may become unstable to plasma instabilities as the solar wind travels outward. For example, the trajectory of both the quiescent and non-quiescent solar wind in Figure \ref{fig:brazilpeaks} is towards the parallel and oblique firehose instabilities. If these regions do become unstable as the solar wind moves outward these instabilities would generate fluctuations in the plasma which would eventually destroy the quiescent signatures of these solar wind regions and may serve to homogenize their turbulent and thermal properties. Alternatively, it may be the case that non-quiescent wind overtakes quiescent wind, causing the two populations to merge. This alternative requires further investigation. 


Previous studies of the solar wind evolution such as \citet{Matteini2007} show that the fast solar wind shows a similar trajectory to the trajectories for quiescent and non-quiescent wind shown in Figure \ref{fig:brazilpeaks}, but in that study the fast solar wind evolves through the parameter space on a much larger radial scale (0.3 AU to \>1 AU). \citet{Matteini2007} also observed that there was no obvious trend derivable from the slow solar wind except that the slow solar wind tended to hover around the firehose instabilities. As such, the data presented here may be consistent with observations from the Helios missions which found that slow solar wind tended to approach the firehose instabilities \citep{Marsch2006, Matteini2007}. This may indicate that the wind observed in this study is consistent with that of the slow wind and may indicate that slow solar wind evolves along similar trajectories as fast solar wind, but on a shorter radial scale.

\subsection{Quiescent Region Magnetic Field Origins} \label{subsec:origins}

For the closest radial bins, quiescent regions have higher proton temperature anisotropies $\mathcal{R}$ than non-quiescent wind at the same altitude above the solar surface (Figures \ref{fig:brazil} and \ref{fig:brazilpeaks}). The innermost radial bin is unique because it is both narrow in radial distance (1.7 $R_{\odot}$ wide), and has a high number of quiescent regions per $R_{\odot}$ (44 quiescent regions per 1.7 $R_{\odot}$) compared to the next radial bin (92 quiescent regions per 10 $R_{\odot}$). This bin has a well-defined peak in the $\mathcal{R}$ vs $\beta_{\parallel p}$ space for both the quiescent solar wind as well as the non-quiescent solar wind. This increase in temperature anisotropy is consistent with a solar wind magnetic field that has not undergone as much expansion as non-quiescent wind during its journey to the spacecraft. 

Consider the case of two adjacent magnetic mirrors whose mirror ends are placed into a source of thermal plasma like the solar surface and whose open ends extend into the solar wind. As protons run up the diverging magnetic field lines, they conserve their adiabatic invariant due to gyro motion, and transfer perpendicular kinetic energy into parallel kinetic energy. If these two magnetic mirrors have different mirror ratios, but comparable field strengths at the base, then the protons which are in the magnetic field which remains more compressed will retain more of their perpendicular kinetic energy, this would manifest as higher temperature anisotropies in the higher field streams. It is demonstrated in Figures \ref{fig:histograms}d1,\ref{fig:histograms}d2, and \ref{fig:histograms}d3 that the magnetic field magnitude for quiescent regions is higher than non-quiescent wind in each radial bin. This is consistent with low levels of magnetic field expansion in quiescent regions as field line density is the definition of magnetic field magnitude. Therefore we suggest that quiescent solar wind regions may exist on magnetic field lines which have undergone less expansion, compared to non-quiescent solar wind, at distances up to a few tens of solar radii ($R_\odot$).

Alternatively, previous studies by \citet{Verniero2020,Verniero2022} have shown that certain cases of quiescent intervals are associated with the presence of ion beams which are in turn associated with ion scale waves. If, for the closest radial bins, quiescent regions are associated with ion scale waves generated by ion beams, wave-particle interactions between protons and proton cyclotron waves \citep{Cranmer2000ApJ,Vech2021,Bowen2022,Bowen2024Nat} may explain the enhanced perpendicular temperature observed by quiescent regions in this radial bin. However, it has not been shown in this study that quiescent regions are associated with ion beams generally and investigations into this possibility are outside the scope of this study.

\section{Conclusion} \label{sec:conclusion}

This study demonstrates that quiescent regions have statistically significant differences in both turbulent and bulk properties and that the evolution of quiescent and non-quiescent wind follows a similar path through instability parameter space. The higher temperature anisotropies and higher magnetic field strengths of quiescent regions in the lowest radial altitude are consistent with magnetic field lines which remain compressed relative to magnetic field lines of non-quiescent solar wind. This study has presented evidence against hypothesis 3 detailed in Section \ref{subsec:introorigins}. A version of hypothesis 3 which involves damping of turbulent fluctuations is unlikely based on energy conservation of the magnetic fluctuations. In quiescent regions, the proton temperature and bulk flow velocity are shown to be depressed or comparable to non-quiescent solar wind, inconsistent with damping of magnetic field fluctuations. This study also estimates a scale height for quiescent regions which characterizes the decay rate of quiescent regions in the solar wind. It is shown that quiescent regions evolve similarly to non-quiescent solar wind but occupy different portions of the $\mathcal{R}$ vs $\beta_{\parallel p}$ parameter space. This may suggest that quiescent region magnetic field lines are different at the respective points of origin for each solar wind population, consistent with hypotheses 1 and 2 discussed in Section \ref{subsec:introorigins}.

While not all possible variations of hypothesis 3 have been explored in this work and has thus not been fully ruled out, we present strong evidence to rule out one scenario and present evidence in favor of hypotheses 1 and 2. Further work is needed to distinguish between hypotheses 1 and 2 as they are variations of hypotheses which have quiescent sources located near the solar surface. Work which distinguishes between hypotheses 1 and 2 will involve dynamics of coronal plasma located very close to the source surface.

\vspace{1cm}

\begin{acknowledgments}
Parker Solar Probe was designed, built, and is now operated by the Johns Hopkins Applied Physics Laboratory as part of NASA’s Living with a Star (LWS) program (contract NNN06AA01C). Support from the LWS management and technical team has played a critical role in the success of the Parker Solar Probe mission. B.M.S. is funded by NASA's Future Investigators in NASA Earth and Space Science and Technology (FINESST) grant (grant number 80NSSC23K1627). J.L.V. acknowledges support from NASA PSP-GI grant 80NSSC23K0208.

Data manipulation was performed using the Numpy \citep{Numpy2020} and Scipy \citep{SciPy2020} Python packages. Figures in this work are produced using the Python package Matplotlib \citep{Matplotlib2007}.
\end{acknowledgments}

\bibliography{references}{}
\bibliographystyle{aasjournal}



\end{document}